\documentclass[12pt]{article}
\usepackage{graphics}
\usepackage{amssymb}
\usepackage{psfrag}

\newcommand{\rf}[1]{(\ref{#1})}
\newcommand{\beq}{\begin{equation}}
\newcommand{\eeq}{\end{equation}}
\newcommand{\bea}{\begin{eqnarray}}
\newcommand{\eea}{\end{eqnarray}}

\newcommand{\e}{\mbox{e}}
\renewcommand{\d}{\mbox{d}}

\newcommand{\lam}{\lambda}
\newcommand{\Lam}{\Lambda}

\renewcommand{\a}{\alpha}

\newcommand{\del}{\delta}
\newcommand{\Del}{\Delta}
\newcommand{\sg}{\sigma}
\newcommand{\ra}{\rangle}
\newcommand{\la}{\langle}
\newcommand{\prt}{\partial}
\newcommand{\mi}{\!-\!}
\newcommand{\equ}{\!=\!}
\newcommand{\pl}{\!+\!}

\newcommand{\cD}{\mathcal{D}}

\newcommand{\cM}{{\cal M}}

\newcommand{\cT}{\mathcal{T}}

\def\R{\, R\!\!\!\!\!\! I\;{}}

\newcommand{\Nof}{N_{14}}

\newcommand{\Ntt}{N_{23}}

\begin{document}
{\normalsize \hfill SPIN-06/16}\\
\vspace{-1.5cm}
{\normalsize \hfill ITP-UU-06/19}\\
${}$\\

\begin{center}
\vspace{24pt}
{ \Large \bf Quantum Gravity, or \\
\vspace{6pt} 
The Art of Building Spacetime}\footnote{Contribution to the book ``
Approaches to Quantum
Gravity'', edited by D. Oriti, to appear at Cambridge University Press.}

\vspace{30pt}

{\sl J. Ambj\o rn}$\,^{a,c}$
{\sl J. Jurkiewicz}$\,^{b}$,
and {\sl R. Loll}$\,^{c}$

\vspace{24pt}
{\footnotesize

$^a$~The Niels Bohr Institute, Copenhagen University\\
Blegdamsvej 17, DK-2100 Copenhagen \O , Denmark.\\
{ email: ambjorn@nbi.dk}\\

\vspace{10pt}

$^b$~Institute of Physics, Jagellonian University,\\
Reymonta 4, PL 30-059 Krakow, Poland.\\
{ email: jurkiewi@thrisc.if.uj.edu.pl}\\

\vspace{10pt}

$^c$~Institute for Theoretical Physics, Utrecht University, \\
Leuvenlaan 4, NL-3584 CE Utrecht, The Netherlands.\\
{ email:  j.ambjorn@phys.uu.nl, r.loll@phys.uu.nl}\\

\vspace{10pt}

}
\vspace{48pt}

\end{center}

%\addtolength{\baselineskip}{0.20\baselineskip}
%\vspace{2cm}

\begin{center}
{\bf Abstract}
\end{center}

The method of four-dimensional Causal Dynamical Triangulations provides
a background-independent definition of the sum over geometries
in quantum gravity, in the presence of a positive cosmological constant.
We present the evidence accumulated to date
that a macroscopic four-dimensional world can emerge from this theory
dynamically. Using computer simulations we observe in the Euclidean
sector a universe whose scale factor exhibits the same dynamics as
that of the simplest mini-superspace models in quantum cosmology, 
with the distinction that in the case of causal dynamical triangulations
the effective action for the scale factor is not put in by hand but obtained 
by integrating out {\it in the quantum theory} the full set of dynamical 
degrees of freedom except for the scale factor itself. 

\vspace{12pt}
\noindent

%\vfill

\newpage

\section{Introduction}\label{intro}

What is more natural than constructing space from elementary
geometric building blocks? It is not as easy as one might think,
based on our intuition of playing with Lego blocks
in three-dimensional space. Imagine the building blocks are $d$-dimensional
flat simplices all of whose side lengths are $a$, and let $d >2$. 
The problem is that if we 
glue such blocks together carelessly we will {\it 
with probability one} create a space of no extension,
in which it is possible
to get from one vertex to any other in a few steps, moving along
the one-dimensional edges of the simplicial manifold we have created. 
We can also say that the space has an extension
which remains at the `cut-off' scale $a$. 
Our intuition coming from playing 
with Lego blocks is misleading here because it 
presupposes that the building blocks are 
embedded geometrically faithfully in Euclidean $\R^{\; 3}$, which is not
the case for the intrinsic geometric construction of a simplicial space.

By contrast, let us now be more careful in our construction
work by assigning to a simplicial space $\cT$ -- which we will interpret
as a (Euclidean) spacetime -- the weight $e^{-S(\cT)}$,
where $S(\cT)$ denotes the Einstein action associated with the piecewise 
linear geometry uniquely defined by our 
construction\footnote{There exists a natural, coordinate-independent 
definition of the Einstein action for piecewise 
linear geometries called the Regge action.}. 
As long as the (bare) gravitational coupling constant $G_N$ is large, 
we have the same situation as before. However, upon lowering $G_N$ we will
eventually encounter a phase transition beyond which 
the geometry is no longer crumpled into a tiny ball,
but maximally extended. Such a geometry is made out of
effectively one-dimensional 
filaments\footnote{The $d$-dimensional building blocks are arranged 
such that $(d-1)$ ``transverse" dimensions have a size of only a few
lattice spacings.} which can branch out, 
and are therefore called {\it branched
polymers} or {\it trees} \cite{aj1,ajjk}. 
The transition separating the two phases \cite{bielefeld,simon}
is of first order, which implies that there is no smooth change between 
the two pathological types of minimally or maximally extended ``universes".

In order for the sum over geometries to produce a quantum theory of 
gravity in which classical geometry is reproduced in a suitable limit, 
we therefore need a different principle for selecting the geometries 
to be included in this sum. Below we will introduce such a principle: 
our prescription will be to sum over a class of (Euclidean) geometries which 
are in one-to-one correspondence
with Lorentzian, {\it causal} geometries. At the discretized level, where 
we use a specific set of building blocks and gluing rules to constructively 
define the path integral, we
call these geometries {\it causal dynamical triangulations} (CDT) 
\cite{al,ajl1,ajl4d,ajl3d}.

Before discussing CDT in more detail let us comment on
the nature of the geometries contributing to the path integral.    
It is important to emphasize that in a {\it quantum} theory of gravity a 
given spacetime geometry as such has no immediate physical meaning. 
The situation is really the same as
in ordinary quantum field theory or even quantum mechanics, where 
individual field configurations $\phi(x,t)$ or particle paths $x(t)$ are
{\it not} observable. Only certain expectation values related to the fields 
or paths can be observed in experiments. This does not mean there
cannot exist limits in which it is appropriate to talk about a 
particular field configuration or
the path of a particle in an approximate sense.
In the case of our actual universe, down to the smallest distances that
have been probed experimentally, it certainly does seem adequate
to talk about a fixed classical spacetime geometry. Nevertheless, at
sufficiently small distances it will no longer make sense to 
ask classical questions about spacetime, at least 
if we are to believe in the principles of 
conventional quantum theory.

By way of illustration let us discuss the situation for the ordinary harmonic 
oscillator (or the free particle) and consider the path integral
from $(x_1,t_1)$ to $(x_2,t_2)$. Precisely for the harmonic oscillator 
(or the free particle) the decomposition
\beq\label{1.1}
x(t) = x_{cl}(t) + y(t),~~~~ y(t_1)=y(t_2)=0,
\eeq
leads to an exact factorization of the path integral, because
the action satisfies
\beq\label{1.1a}
S(x)=S(x_{cl})+S(y).
\eeq
This implies that the classical path $x_{cl}(t)$ contributes to the path integral 
with the classical action, and $y(t)$ 
with quantum fluctuations independent of this classical part. 
Taking the classical trajectory to be macroscopic one obtains the picture of
a macroscopic path dressed with small quantum fluctuations; small
because they are independent of the classical motion. An explicit Euclidean 
calculation yields the result
\beq\label{1.2}
\left\la \int_0^T \d t\; y^2(t)\right\ra = 
\frac{\hbar}{2m\omega^2}\;(\omega T\tanh^{-1} \omega T-1)
\eeq
as a function of the oscillator frequency $\omega$ and mass $m$.
Let us now consider a situation where we have chosen the ``system size", 
i.e. $x_{cl}(t)$, to be macroscopic. According to \rf{1.2}, the quantum 
fluctuations around 
this path can then be considered small since $\hbar$ is small.  

This is more or less the picture we envisage for our present-day universe
in quantum gravity: the universe is of macroscopic size, governed
by the classical equations of motion (the analogue of choosing
``by hand" $(x_1,t_1)$ and $(x_2,t_2)$ to be macroscopic in the example above),
and the small quantum fluctuations are dictated
by the gravitational coupling constant (times $\hbar/c^3$). 

A given configuration $x(t)$ in the path integral for the quantum-mechanical
particle is (with probability one) a continuous, nowhere differentiable path, 
which moreover is fractal with Hausdorff dimension two,
as we know from the rigorous construction of the Wiener
measure on the set of parametrized paths. In the case of 
quantum gravity we do not have a similar mathematically rigorously defined
measure on the space of geometries, but it is natural to expect that 
{\it if} it exists, a typical geometry  in the path integral
will be continuous, but nowhere differentiable.
By analogy, the piecewise linear geometries seem a good choice
if we want to approximate the gravitational path integral
by a set of geometries and subsequently 
take a limit where the approximation (the cut-off) is removed.
Moreover, such simplicial manifolds possess a natural, geometric and 
coordinate-independent implementation of
the Einstein-Hilbert action. With all local curvature degrees of freedom
present (albeit in a discretized fashion), we also expect them
to be suitably ``dense'' in the set of all continuous geometries.

The spirit is very much that of the standard lattice formulation of
quantum field theory where (flat) spacetime is approximated by 
a hypercubic lattice. The ultraviolet cut-off in such field theories 
is given by the lattice spacing, i.e. the length of all one-dimensional
lattice edges. We can in a similar and simple manner introduce 
a {\it diffeomorphism-invariant} cut-off in the sum over the 
piecewise linear geometries by restricting it to
the {\it building blocks} mentioned earlier. 
A natural building block for a $d$-dimensional 
spacetime is a $d$-dimensional equilateral simplex with 
side-length $a$, and the path integral is approximated by performing 
the sum over all geometries (of fixed topology\footnote{In
classical General Relativity there is no motivation to 
consider spacetimes whose spatial topology changes in time,
since their Lorentzian structure is necessarily singular.
There is an interesting and long-standing 
discussion about whether one should include topology changes in a 
{\it quantum} theory of gravity.  
However, even in the case of two-dimensional Euclidean quantum
gravity, where the classification of topology changes is simple, the 
summation over topologies has never been defined  
non-perturbatively in a satisfactory way, despite many 
attempts, in particular, in so-called non-critical string theory. 
(However, see \cite{cdt-topo} for how one may improve the
convergence of the sum in two-dimensional {\it Lorentzian}
quantum gravity by invoking not just the topological, but the
causal, geometric structure of spacetime.) The situation
becomes worse in higher dimensions.
For instance, four-dimensional topologies
are not classifiable, so what does it mean to sum over them
in the path integral? The problem -- even in dimension two -- is
that there are many more geometries of complicated topology than 
there are of simple topology, with the consequence that 
any sum over geometries will be (i)
completely dominated by these complicated topologies, and (ii) plainly 
divergent in a way which (until now) has made it impossible to define 
the theory non-perturbatively in an unambiguous and physically satisfactory  
manner. In higher dimensions these problems are totally out of control.}) 
which can be obtained by gluing 
such building blocks together, each geometry weighted appropriately (for
example, by 
$e^{-S}$, where $S$ is the Einstein-Hilbert action). Afterwards we take 
the limit $a \to 0$. For a particular choice of the bare, dimensionless 
coupling constants one may be able to obtain a continuum limit,
and thus extract a continuum theory. For other values, if the sum exists
at all (possibly after renormalization), one will merely 
obtain a sum which has no continuum interpretation. This situation
is precisely the same as encountered in ordinary lattice field 
theory in flat spacetime.

As mentioned earlier it has up to now not been possible to 
define constructively a Euclidean path integral for gravity 
in four dimensions by following the philosophy just outlined. One 
simply has not succeeded in identifying a continuum limit of the 
(unrestricted) sum over Euclidean building blocks. Among the 
reasons that have been advanced to explain 
this failure, it is clear that the {\it entropy} of the various geometries
plays an important role. We have already pointed out that the crumpled 
geometries of no extension dominate the space of all continuous geometries
whenever the dimension of spacetime is larger than two.
There is nothing wrong with this a priori; the path integral of any quantum 
field theory is dominated completely by wild UV-field fluctuations.
However, in the case of {\it renormalizable} quantum field theories there 
exists a well-defined limiting procedure which allows one to extract 
``continuum'' physics by fine-tuning the bare coupling constants of the 
theory. An analogous procedure in Euclidean quantum gravity still has
not been found, and adding (bosonic) matter does not improve the 
situation. Instead, note that the Einstein-Hilbert action has a unique feature, 
namely, it is unbounded from below.
The transition between the crumpled and the branched-polymer geometries 
can be seen as a transition from 
a phase where the entropy of configurations dominates over the 
action to a phase where the unboundedness of the Euclidean action 
becomes dominant\footnote{Although the action is not unbounded 
below in the regularized theory, this feature of the continuum
action nevertheless manifests itself in the limit as the (discretized) volume of 
spacetime is increased, eventually 
leading to the above-mentioned phase transition
at a particular value of the bare gravitational coupling 
constant.
Remarkably, a related phenomenon occurs in bosonic string 
theory. If the world-sheet theory is regularized non-perturbatively 
in terms of triangulations (with each two-dimensional world-sheet
glued from fundamental simplicial building blocks),
the tachyonic sickness of the theory manifests itself in the
form of surfaces degenerating into branched polymers \cite{ad}.}.
The impossibility of finding a continuum limit may be seen as the 
impossibility of balancing the
entropy of configurations against the action. 
We need another guiding principle for selecting 
Euclidean geometries in the path integral in order to obtain a
continuum limit, and it is such a principle we turn to next.

\section{Defining CDT}

It has been suggested that the signature of spacetime
may be explained from a dynamical principle \cite{jeff}.
Being somewhat less ambitious, we will assume it has 
Lorentzian signature and accordingly change our
perspective from the Euclidean formulation of the path
integral discussed in the previous section to a Lorentzian formulation,
motivated by the uncontroversial fact that our universe has
three space and one time dimension.
A specific rotation to Euclidean signature introduced below
will be needed in our set-up as a merely technical tool
to perform certain sums over geometries.
Unlike in flat spacetime there are no general
theorems which would allow us to relate the Euclidean and Lorentzian 
quantum field theories when dealing with quantum gravity.

Consider now a connected space-like hypersurface in spacetime.
Any classical evolution in general relativity will leave the
topology of this hypersurface unchanged, since otherwise
spacetime would contain regions where the metric is degenerate.
However, as long as we do not have a 
consistent theory of quantum gravity we do not know whether
such degenerate configurations should be included in the path integral.
We have already argued that the inclusion of arbitrary 
spacetime topologies leads to a path
integral that has little chance of making sense.
One might still consider a situation where the overall topology of 
spacetime is fixed, but where one allows ``baby universes"
to branch off from the main universe, without permitting them
to rejoin it and thus form ``handles". Apart from being
a rather artificial constraint on geometry, such a construction
is unlikely to be compatible with unitarity.
We will in the following take a conservative point of view and only sum 
over geometries (with Lorentzian signature) which permit a 
foliation in (proper) time and are causally well-behaved in the sense that
no topology changes are allowed as a function of time.
In the context of a formal continuum path integral for gravity,
similar ideas have earlier been advanced in \cite{teitelboim}.

Of the diffeomorphism-invariant quantities one can consider
in the quantum theory, we have chosen a particular {\it proper-time 
propagator}, which can be defined constructively in a
transparent way. 
We are thus interested in defining 
the path integral
\beq\label{2.3}
G(g(0),g(T);T)= \int_{g(0)}^{g(T)} \cD g\; \e^{iS[g]}
\eeq
over Lorentzian geometries on a manifold $\cM$ with 
topology $\Sigma\times [0,1]$, where $\Sigma$ is a compact, 
connected three-dimensional manifold. The geometries 
included in the path integral will be such that 
the induced boundary three-geometries $g(0)$ and $g(T)$ are space-like
and separated by a time-like geodesic distance $T$, with
$T$ an external (diffeomorphism-invariant) parameter.    

We now turn to the constructive definition of this object in terms
of building blocks. 
The discretized analogue of an infinitesimal proper-time
``sandwich" in the continuum will be a finite sandwich 
of thickness $\Delta t=1$ (measured in ``building block units'' $a$)
of topology $\Sigma\times [0,1]$ 
consisting of a single layer of
four-simplices. This layer has two spacelike boundaries,
corresponding to two slices of constant (integer) ``proper
time" $t$ which are one unit apart. They form two
three-dimensional piecewise flat manifolds of topology $\Sigma$
and consist of purely spacelike tetrahedra. By construction,  
the sandwich interior contains no vertices, so that any one
of the four-simplices shares $k$ of its vertices with the initial
spatial slice and $5-k$ of them with the final spatial slice,
where $1\leq k\leq 4$. To obtain extended
spacetimes, one glues together sandwiches pairwise along
their matching three-dimensional boundary geometries. 
We choose each four-simplex to have time-like
links of length-squared $a_t^2$ and space-like links
of length-squared $a_s^2$, with all of the latter located in spatial
slices of constant integer-$t$. 

\begin{figure}[t]
\centerline{\scalebox{0.6}{\rotatebox{0}{\includegraphics{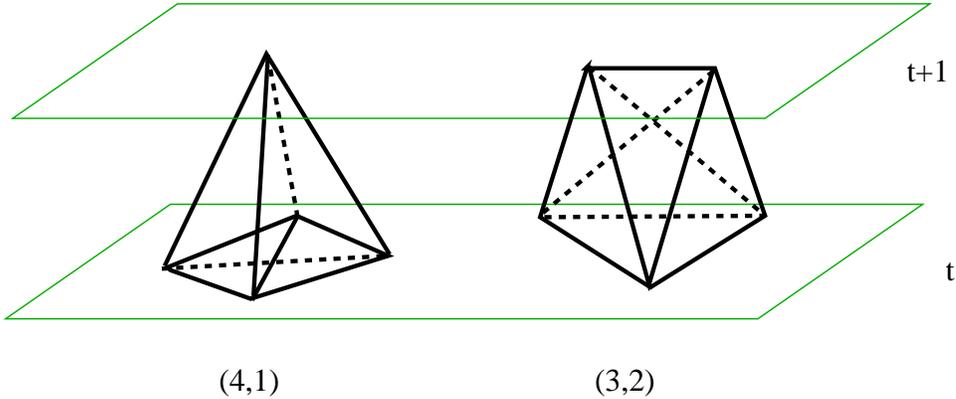}}}}
\caption[phased]{{\small
The two fundamental building blocks of causal dynamically triangulated gravity.
The flat four-simplex of type (4,1) on the left 
has four of its vertices at time $t$ and
one at time $t\pl 1$, and analogously for the (3,2)-simplex on the right. 
The ``gap" between two consecutive 
spatial slices of constant integer time is filled by copies
of these simplicial building blocks and their time-reversed counterparts,
the (1,4)- and the (2,3)-simplices.
}}
\label{fig1}
\end{figure}

Each spatial tetrahedron at time $t$ is therefore
shared by two four-simplices (said to be of type (1,4) and (4,1)) 
whose fifth vertex lies in the neighbouring
slice of constant time $t-1$ and $t+1$ respectively. In addition
we need four-simplices of type (2,3) and (3,2) which share one link and one 
triangle with two adjacent spatial slices, as illustrated in
Fig.\ \ref{fig1} (see \cite{ajl4d} for details). 
The integer-valued proper time $t$ can be extended in a natural
way to the interiors of the four-simplices, leading to a global foliation
of any causal dynamically triangulated spacetime into piecewise
flat (generalized) triangulations for any constant real value of $t$
\cite{dl}. Inside each building block this time coincides
with the proper time of Minkowski space. 
Moreover, it can be seen that in the piecewise linear geometries
the mid-points of all
spatial tetrahedra at constant time $t$ are separated a 
fixed time-like geodesic distance 1 (in units of $a_t$) 
from the neighbouring hypersurfaces at $t- 1$ and $t+ 1$. 
It is in this sense that the ``link distance" $t$, i.e. counting 
future-oriented time-like links between spatial slices
is a discretized analogue of their proper-time distance.

Let us furthermore assume that the two possible link lengths
are related by
\beq\label{3.1a}
a_t^2 = -\a a_s^2.
\eeq
All choices $\a > 0$ correspond to Lorentzian and all choices $\a < -7/12$
to Euclidean signature, and a Euclideanization of geometry is obtained
by a suitable analytic continuation in $\alpha$ (see \cite{ajl4d} for 
a detailed discussion of this ``Wick rotation" where one finds
$S_E (-\a) = i S_L (\a )$ for $\a > 7/12$).

Setting $\a = -1$ leads to a particularly 
simple expression for the (Euclidean) Einstein-Hilbert action of a given 
triangulation $\cT$ (since all four-simplices
are then identical geometrically), namely, 
\beq\label{3.2a}
S_E(\cT) = -k_0 N_0(\cT) + k_4 N_4(\cT),
\eeq
with $N_i(\cT)$ denoting the number of $i$-dimensional simplices in
$T$. In \rf{3.2a}, $k_0$ is proportional 
to the inverse (bare) gravitational coupling 
constant, $k_0 \sim 1/G_N$, while $k_4$ is a 
linear combination of the cosmological and inverse gravitational 
coupling constants. The action \rf{3.2a} is calculated 
from Regge's prescription for piecewise linear geometries.
If we take $\a \neq -1$ the Euclidean 
four-simplices of type (1,4) and type (2,3)
will be different and appear with different weights in the 
Einstein-Hilbert action \cite{ajl4d}. For our present purposes 
it is convenient to use the equivalent parametrization
\beq\label{3.3a}
S_E(\cT) = -k_0 N_0(\cT) + k_4 N_4(\cT) + \Del (2\Nof(\cT)+\Ntt(\cT)),
\eeq
where $\Nof(\cT)$ and $\Ntt(\cT)$ denote the combined numbers
in $\cT$ of four-simplices
of types $(1,4)$ and $(4,1)$, and of types
$(2,3)$ and $(3,2)$, respectively.
The explicit map between the parameter $\Del$ in eq.\ \rf{3.3a} and $\a$ 
can be readily worked out \cite{ajl5}. For 
the simulations reported here we have used $\Del$ in the range 0.4--0.6.

The (Euclidean) discretized analogue of the continuum proper-time propagator 
\rf{2.3} is defined by
\beq\label{3.4}
G_{k_0,k_4,\Del}( \cT^{(3)}(0),\cT^{(3)}(T),T) = 
\sum_{\cT\in\cT_T} \frac{1}{C_{\cT}} \; e^{-S_E(\cT)}. 
\eeq
where the summation is over the set $\cT_T$ of all four-dimensional 
triangulations of topology $\Sigma^3\times [0,1]$ (which we 
in the following always choose to be $S^3$) and $T$
proper-time steps, whose spatial 
boundary geometries at proper times $0$ and $T$ are
$\cT^{(3)}(0)$ and $\cT^{(3)}(T)$.
The order of the automorphism group of the graph $\cT$ is denoted 
by $C_{\cT}$. The propagator can be 
related to the quantum Hamiltonian conjugate to $t$, and
in turn to the transfer matrix of the (Euclidean) statistical theory
\cite{ajl4d}. 

It is important to emphasize again that we rotate each configuration
to a Euclidean ``spacetime" simply in order 
to perform the summation in the path integral, and that this is made possible 
by the piecewise linear structure of our geometry and the existence of 
a proper-time foliation. Viewed from an inherently Euclidean perspective 
there would be no motivation to restrict the sum over geometries to 
``causal'' geometries of the kind constructed above.
We also want to stress that the use of piecewise linear geometries
has allowed us to write down a (regularized) version of \rf{2.3}
using only geometries, not metrics (which are of course not 
diffeomorphism-invariant), and finally that the use of building 
blocks has enabled the introduction of a diffeomorphism-invariant 
cut-off (the lattice link length $a$).

\section{Numerical analysis of the model}

While it may be difficult to find an explicit analytic
expression for the full propagator \rf{3.4} of the four-dimensional theory,
Monte Carlo simulations are readily available for its analysis, employing standard
techniques from Euclidean dynamically triangulated quantum gravity
\cite{leshouches}.  Ideally one would like to keep the
{\it renormalized}\ \footnote{For the relation between the bare
(dimensionless) cosmological constant $k_4$ and the renormalized
cosmological constant $\Lam$ see \cite{aj1}.}
cosmological constant $\Lambda$ fixed in the simulation, in which case
the presence of the cosmological term $\Lam \int \sqrt{g}$ in the action would
imply that the four-volume $V_4$ fluctuated around
$\la V_4 \ra \sim \Lam^{-1}$. However,
for simulation-technical reasons one fixes instead
the number $N_4$ of four-simplices (or\footnote{For fixed 
$\a$ (or $\Delta$) one has
$\la \Nof \ra \propto \la \Ntt \ra \propto \la N_4 \ra$. $V_4$ is
given as (see \cite{ajl4d} for details):
$V_4 = a_s^4(\Nof\sqrt{8\a+3} + \Ntt\sqrt{12\a +7}).$
We set $a_s=1$.}  the four-volume $V_4$)
from the outset, working effectively
with a cosmological constant $\Lam \sim V_4^{-1}$.

\subsection{The global dimension of spacetime}

\begin{figure}[t]
%\vspace{-3cm}
\centerline{\scalebox{1.0}{\rotatebox{0}{\includegraphics{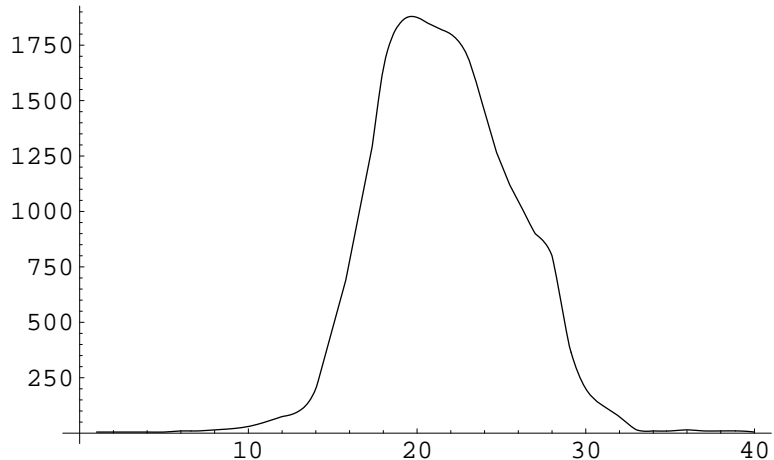}}}}
%\vspace{-3.5cm}
\caption[phased]{{\small
Snapshot of a ``typical universe'' consisting of approximately
91000 four-simplices as it appears in
the Monte Carlo simulations at a given ``computer time''.
We plot the three-volume at each integer step in proper time,
for a total time extent of $T=40$, in units where $a_s=1$.}}
\label{fig0}
\end{figure}

A ``snapshot", by which we mean the distribution of three-volumes
as a function of the proper time $0\leq t\leq T$ for
a spacetime configuration randomly picked from the
Monte Carlo-generated geometric ensemble, is shown in Fig.\ \ref{fig0}.
One observes a``stalk" of essentially no spatial extension (with spatial 
volumes close to  the minimal triangulation of $S^3$ consisting of
five tetrahedra) expanding into a universe of genuine
``macroscopic" spatial volumes, which after a certain time $\tau \leq T$
contracts again to a state of minimal spatial extension.
As we emphasized earlier, a single such configuration is
unphysical, and therefore not observable. However, a more
systematic analysis reveals that fluctuations around an overall
``shape" similar to the one of  Fig.\ \ref{fig0}
are relatively small, suggesting the existence of a {\it background
geometry} with relatively small quantum fluctuations superimposed.
This is precisely the scenario advocated in Sec.\ 1 and is rather
remarkable, given that our formalism
is background-independent. Our first major goal is to verify
quantitatively that we are indeed dealing with an approximate
four-dimensional background geometry \cite{ajl-prl,spectral},
and secondly to determine the effective action responsible for the
observed large-scale features of this background geometry \cite{semi,ajl5} .

Important information is contained in how the expectation values of
the volume $V_3$ of
spatial slices and the total time extent $\tau$ (the proper-time interval
during which the spatial volumes $V_3 \gg 1$)
of the observed universe behave as the total spacetime volume $V_4$
is varied.
We find that to good approximation the spatially extended parts of
the spacetimes for various four-volumes $V_4$ can be
mapped onto each other by rescaling the spatial volumes
and the proper times according to
\beq\label{3.5}
V_3 \to V_3/V_4^{3/4},~~~~\tau \to \tau/V_4^{1/4}.
\eeq
To quantify this we studied the so-called volume-volume
correlator 
\beq\label{v-v}
\la V_3(0)V_3(\del)\ra = \frac{1}{t^2} \sum_{j=1}^{t}
\la V_3(j)V_3(j+\del) \ra
\eeq
for pairs of spatial slices an integer proper-time distance
$\delta$ apart. Fig.~\ref{fig3} shows the volume-volume
correlator for five different spacetime volumes $V_4$, using the rescaling
\rf{3.5}\footnote{In \rf{v-v} we use discrete units such that
successive spatial slices are separated by 1.
For convenience we periodically identify $\cT^{(3)}(T)=\cT^{(3)}(0)$
and sum over all possible three-geometries $\cT^{(3)}(0)$, 
rather than working with fixed boundary conditions. 
In this way \rf{v-v} becomes a convenient 
translation-invariant measure of the spatial and temporal extensions of
the universe (see \cite{ajl3d} for a detailed discussion).},
\begin{figure}[ht]
\vspace{-3cm}
\psfrag{d}{\bf{\Large $x$}}
\psfrag{VV}{\Large\bf $c_{ N_4}(x)$}
\centerline{\scalebox{0.7}{\rotatebox{0}{\includegraphics{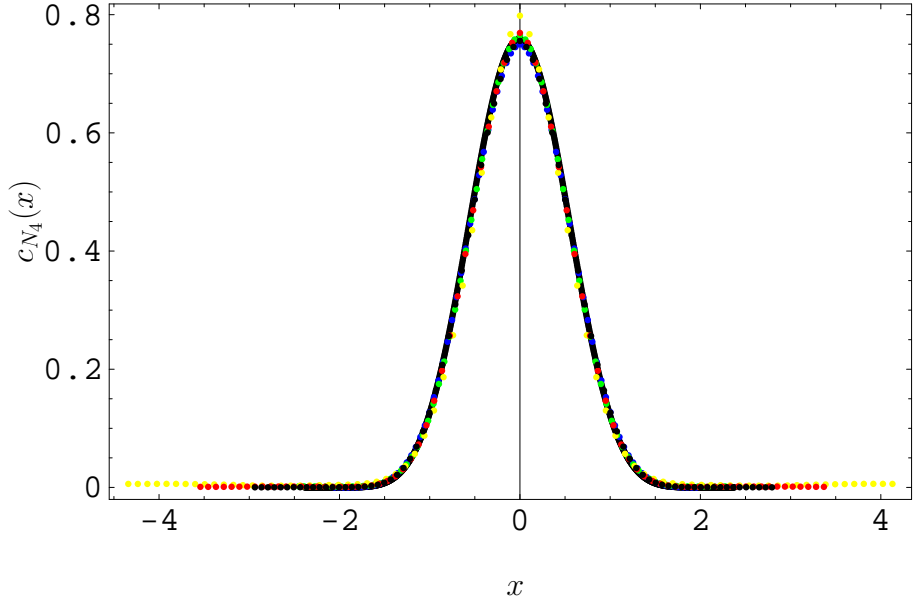}}}}
\vspace{-4.5cm}
\caption[phased]{{The scaling of the volume-volume correlator,
as function of the rescaled time variable 
$x\equ \delta/( N_4)^{1/4}$.
Data points come from system sizes $ N_4\equ$ 22500, 45000, 91000, 
181000 and 362000 at
$\kappa_0\equ 2.2$, $\Delta\equ 0.6$ and $T\equ 80$.
}}
\label{fig2}
\end{figure}
and exhibiting that it is almost perfect. An error estimate
yields $d=4\pm 0.2$ for the large-scale dimension of the universe \cite{ajl5}.

Another way of obtaining an effective dimension of the
nonperturbative ground state, its so-called {\it spectral dimension} $D_S$,
comes from studying a diffusion
process on the underlying geometric ensemble. On a $d$-dimensional 
manifold with a fixed,
smooth Riemannian metric $g_{ab}(\xi)$, the diffusion equation has the form
\beq\label{ja2}
\frac{\prt}{\prt \sg} \, K_g(\xi,\xi_0;\sg) = \Del_g K_g (\xi,\xi_0;\sg),
\eeq
where $\sg$ is a fictitious diffusion time, $\Del_g$ the Laplace
operator of the metric $g_{ab}(\xi)$ and $K_g(\xi,\xi_0;\sg)$
the probability density of diffusion from point $\xi_0$ to point $\xi$ in
diffusion time $\sg$. We will consider diffusion processes which 
initially are peaked at some point $\xi_0$, so that
\beq\label{ja3}
K_g(\xi,\xi_0;\sg\equ 0) = \frac{1}{\sqrt{\det g(\xi)}}\, \del^d(\xi-\xi_0).
\eeq
For the special case of a flat Euclidean metric, we have
\beq\label{flat1}
K_g(\xi,\xi_0;\sg) = \frac{\e^{-d_g^2(\xi,\xi_0)/4\sg}}{(4\pi \sg)^{d/2}},
\qquad g_{ab}(\xi)\equ \delta_{ab},
\eeq
where $d_g$ denotes the distance function associated with the metric $g$.

A quantity which is easier to measure in numerical simulations is the {\it average 
return probability} $P_{g}(\sg)$, defined by 
\beq\label{ja5}
P_{g}(\sg) := \frac{1}{V} \int  
d^d\xi \sqrt{\det g(\xi)} \; K_g(\xi,\xi;\sg), 
\eeq
where $V$ is the spacetime volume $V\equ\int d^d\xi \sqrt{\det g(\xi)}$. 
For an infinite flat space, 
we have $P_g(\sg)\equ  1/(4\pi\sg )^{d/2}$ and thus can
extract the dimension $d$ by taking the logarithmic derivative
\beq\label{ja5a}
-2\ \frac{d \log P_g(\sg)}{d\log \sg} = d,
\eeq
independent of $\sg$. For nonflat spaces and/or finite volume $V$,
one can still use
eq.\ \rf{ja5a} to extract the dimension, but there will be
correction terms (see \cite{ajl5} for a detailed discussion).

In applying this set-up to four-dimensional 
quantum gravity in a path integral formulation,
we are interested in measuring the expectation value of the average return
probability $P_g(\sg)$. Since $P_g(\sg)$ defined according to \rf{ja5}
is invariant under reparametrizations, it
makes sense to take its quantum average over all geometries of
a given spacetime volume $V_4$,
\beq\label{ja7}
P_{V_4}(\sg) =
\frac{1}{\tilde Z_E(V_4)} \int\!\! \cD [g_{ab}] \; e^{-\tilde{S}_E(g_{ab})}
\del(\int d^4x \sqrt{\det g}-V_4) \, P_g(\sg),
\eeq
where $\tilde Z_E(V_4)$ is the quantum gravity partition function
for spacetimes with constant four-volume $V_4$.

Our next task is to define a diffusion process on the class of metric spaces
under consideration, the piecewise linear structures
defined by the causal triangulations $\cT$.
We start from an initial probability distribution 
\beq
K_\cT(i,i_0;\sg\equ 0) = \delta_{i,i_0},
\eeq
which vanishes everywhere except at a randomly chosen (4,1)-simplex $i_0$,
and define the diffusion process by the evolution rule
\beq
K_\cT(j,i_0;\sg+1) = \frac{1}{5}\sum_{k\to j} K_\cT(k,i_0;\sg),
\label{evo4d}
\eeq
where the diffusion time $\sigma$ now advances in discrete integer steps.
These equations are the simplicial analogues of \rf{ja3} and \rf{ja2},
$k\to j$ denoting the five nearest neighbours of the four-simplex $j$.
In this process, the total probability
\beq
\sum_j K_\cT(j,i_0;\sg) =1
\eeq
is conserved. The probability to return to the simplex $i_0$ is then
defined as $P_\cT(i_0;\sg)\equ  K_\cT(i_0,i_0;\sg)$ and its
quantum average as
\beq\label{ja7a}
P_{N_4}(\sg)= \frac{1}{\tilde Z_E(N_4)} \sum_{\cT_{N_4}} 
e^{-\tilde{S}_E(\cT_{N_4})}\;
\frac{1}{N_4}\sum_{i_0 \in \cT_{N_4}} K_{\cT_{N_4}}(i_0,i_0;\sg),
\eeq
where $\cT_{N_4}$ denotes a triangulation with $N_4$ four-simplices,
and $\tilde{S}_E(\cT_{N_4})$ and $\tilde Z_E(N_4)$ are the obvious
simplicial analogues of the continuum quantities in eq.\ \rf{ja7}.

We can extract the value of the spectral dimension $D_S$ 
by measuring the logarithmic derivative as in \rf{ja5a} above, 
that is,
\beq\label{ja1}
D_S(\sg) = -2 \;\frac{d\log P_{N_4}(\sg)}{d\log \sg},
\eeq
as long
as the diffusion time is not much larger than $N_4^{2/D_S}$.
\begin{figure}[t]
%\vspace{-3cm}
\psfrag{X}{{$\sg$}}
\psfrag{Y}{{ $D_S$}}
\centerline{\scalebox{1.2}{\rotatebox{0}{\includegraphics{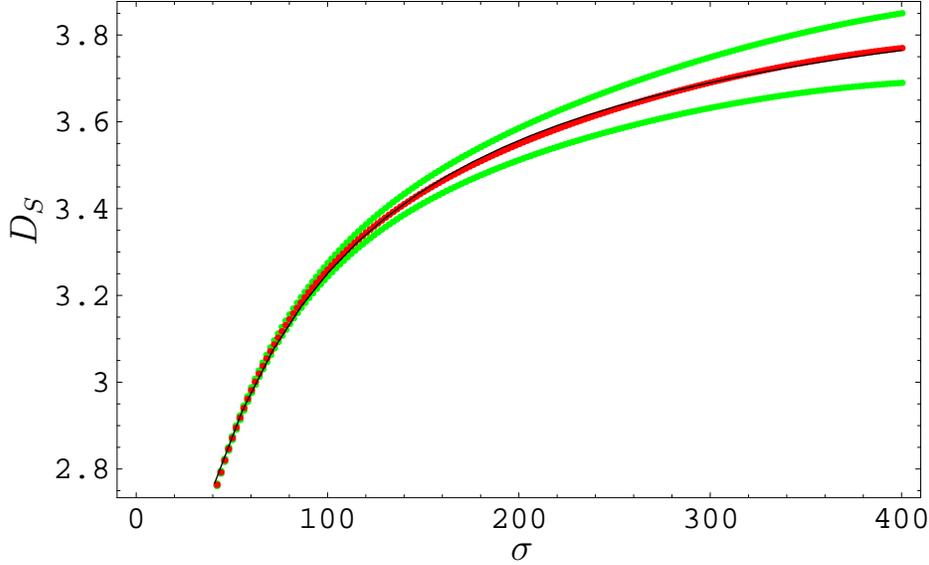}}}~~~~
~~~}
%\vspace{-1cm}
\caption[phased]{{ The spectral dimension $D_S$ of the universe as
function of the diffusion time $\sg$, measured for
$\kappa_0\equ 2.2$, $\Delta\equ 0.6$ and $t\equ 80$, and a spacetime volume
$N_4\equ 181$k. The averaged measurements lie along the central curve,
together with a superimposed best fit 
$D_S(\sg) = 4.02\mi 119/(54\pl\sg)$ (thin black curve). The two outer
curves represent error bars.}}
\label{d4s2.2b4}
\end{figure}
The outcome of the measurements is presented in Fig.\ \ref{d4s2.2b4},
with error bars included. (The two outer curves
represent the envelopes to the tops and bottoms of the error bars.) The
error grows linearly with $\sg$, due to the presence of 
the log$\, \sg$ in \rf{ja1}.

The remarkable feature of the curve $D_S(\sg)$ is its slow approach to the
asymptotic value of $D_S(\sg)$ for large $\sg$. 
The new phenomenon we observe here is a
{\it scale dependence of the spectral dimension}, which has emerged
dynamically \cite{spectral,ajl5}. 

As explained in \cite{spectral}, the best three-parameter fit which
asymptotically approaches a constant is of the form
\beq\label{ja9}
D_S(\sg) =  a -\frac{b}{\sg+c} = 4.02-\frac{119}{54+\sg}.
\eeq
The constants $a$, $b$ and $c$ have been determined by using the 
data range $\sg \in [40,400]$
and the curve shape agrees well with the measurements, as can be seen from 
Fig.\ \ref{d4s2.2b4}.
Integrating \rf{ja9} we obtain
\beq\label{ja8a}
P(\sg) \sim \frac{1}{\sg^{a/2} (1+c/\sg)^{b/2c}},
\eeq
from which we deduce the limiting cases
\beq\label{ja8b}
P(\sg) \sim \left\{
\begin{array}{cl}
\displaystyle{{\sg^{-a/2}}} &~~\mbox{for large $\sg$,}\\
~&~\\
\displaystyle{{\sg^{-(a-b/c)/2}}} &~~ \mbox{for small $\sg$}.
\end{array}
\right.
\eeq
{\it Again we conclude that within measuring accuracy the large-scale dimension 
of spacetime in our model is four. We also note that the short-distance 
spectral dimension seems to be approximately $D_S= 2$, signalling a
highly non-classical behaviour.}

\subsection{The effective action}

\begin{figure}[ht]
\vspace{-3cm}
\psfrag{x}{\bf{\Large $z$}}
\psfrag{T}{\Large\bf $P_{V_3}(z)$}
\centerline{\scalebox{0.6}{\rotatebox{0}{\includegraphics{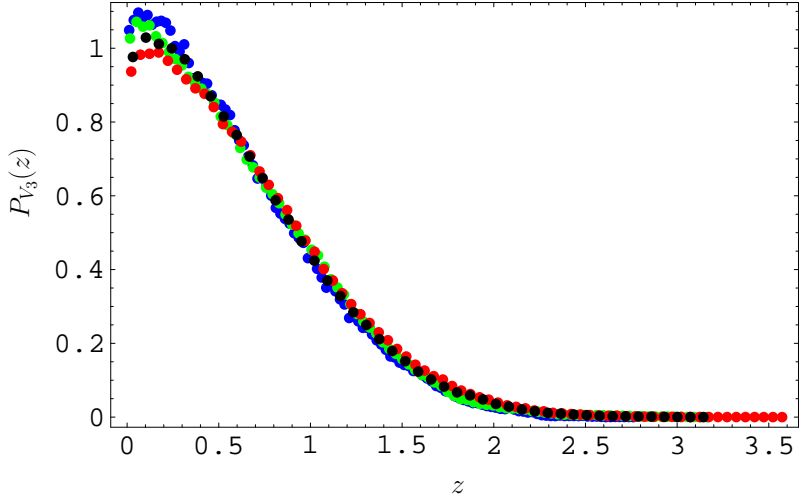}}}}
\vspace{-4.5cm}
\caption[phased]{{\small Distribution $P_{V_3}(z)$ of volume differences
of adjacent spatial slices, for three-volumes $V_3=$ 10.000, 20.000, 
40.000 and 80.000 tetrahedra.}}
\label{fig3}
\end{figure}

{\it Our next goal will be to 
understand the precise analytical form of the volume-volume correlator
\rf{v-v}.} To this end, let us consider the 
distribution of differences in the spatial volumes $V_3$ of 
successive spatial slices 
at proper times $t$ and $t+\del$, 
where $\del$ is infinitesimal, i.e.\ $\del=1$ in lattice 
proper-time units. We have measured the probability 
distribution $P_{V_3}(z)$ of the variable
\beq\label{z}
z= \frac{V_3(t+\del)-V_3(t)}{V_3^{1/2}},~~~~V_3:=V_3(t)+V_3(t+\del).
\eeq
for different values of $V_3$. As shown in 
Fig.\ \ref{fig3} they fall on a common curve.\footnote{Again we have 
applied finite-size scaling techniques, starting out
with an arbitrary power $V_3^{\a}$ in the denominator in \rf{z}, 
and then determining $\a =1/2$
from the principle of maximal overlap of the distributions for various 
$V_3$'s.} Furthermore, the distribution $P_{V_3}(z)$ is fitted very well by 
a Gaussian $e^{-c z^2}$, with a constant $c$ independent of $V_3$.   
From estimating the entropy of spatial geometries, that is, 
the number of such configurations,
one would expect corrections of the form  $V_3^\a$, with $0\leq \a < 1$, 
to the exponent $c \, z^2$ in the distribution $P_{V_3}(z)$. Unfortunately 
it is impossible to measure these corrections directly in a reliable way.
We therefore make a general ansatz for the probability distribution 
for {\it large} $V_3(t)$ as
\beq\label{2.6}
\exp\left[- \frac{c_1}{V_3(t)} \left(\frac{\d V_3(t)}{\d t}\right)^2
-c_2 V_3^{\a}(t)\right],
\eeq
where $0\leq \a <1$, and $c_1$ and $c_2$ are positive constants. 

In this manner, we are led by  ``observation'' to the 
effective action
\beq\label{2.7}
S^{\it eff}_{V_4} = \int_0^T \d t \; 
\left(\frac{c_1}{V_3(t)} \left(\frac{\d V_3(t)}{\d t}\right)^2
+c_2 V_3^{\a}(t) - \lambda V_3(t)\right),   
\eeq
valid for large three-volume $V_3(t)$,
where $\lambda$ is a Lagrange multiplier to be determined 
such that
\beq\label{2.8}
\int_0^T dt \, V_3(t) = V_4.
\eeq
From general scaling of the above action it is clear that
the only chance to obtain the observed scaling law, expressed in terms of
the variable $t/V_4^{1/4}$, is by setting $\a= 1/3$. In addition, to reproduce
the observed stalk for large times $t$ the function $V_3^{1/3}$ has to
be replaced by a function of $V_3$ whose derivative at 0 goes like
$V_3^\nu$, $\nu\geq 0$, for reasons that will become
clear below. A simple
modification, which keeps the large-$V_3$ behaviour intact, is given by
\beq\label{2.9}
V_3^{1/3} \to (1+V_3)^{1/3}-1,
\eeq
but the detailed form is not important. If we
now introduce the (non-negative) {\it scale~factor} $a(t)$ by
\beq\label{2.10}
V_3(t) = a^3(t),
\eeq
we can (after suitable rescaling of $t$ and $a(t)$) 
write the effective action as
\beq\label{2.11}
S^{\it eff}_{V_4} = \frac{1}{G_N} \int_0^T \d t \;
\left( a(t)\left(\frac{\d a(t)}{\d t}\right)^2
+ a(t) - \lam a^3 (t)\right),
\eeq
with the understanding that the linear term should be replaced 
using \rf{2.10} and \rf{2.9} for small $a(t)$.
We emphasize again that we have been led to \rf{2.11}
entirely by  ``observation'' and that one can view
the small-$a(t)$ behaviour implied by \rf{2.9}
as a result of quantum fluctuations.

\subsection{Minisuperspace}

\begin{figure}[ht]
%\vspace{-3cm}
\centerline{\scalebox{0.7}{\rotatebox{0}{\includegraphics{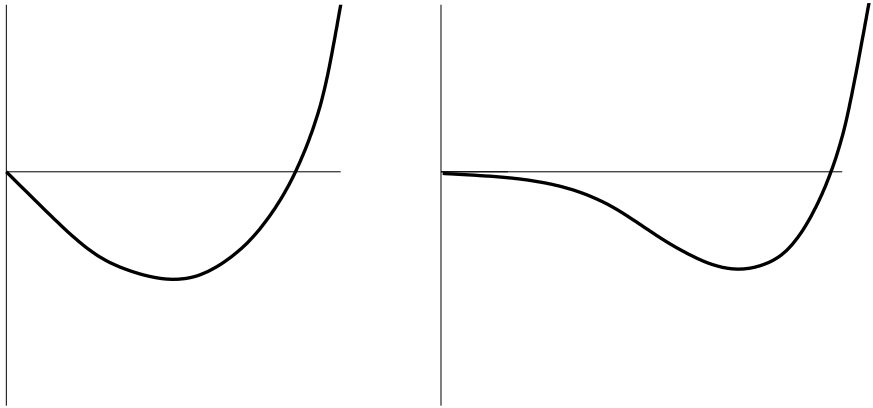}}}}
%\vspace{-3.5cm}
\caption[phased]{{\small The potential $V(a)$ of \rf{3.3}
underlying the standard minisuperspace
dynamics (left)
and the analogous potential in the effective action obtained from the
full quantum gravity model, with small-$a$ modification due to quantum
fluctuations (right).}}
\label{fig4}
\end{figure}

Let us now consider the simplest minisuperspace model for a
closed universe in quantum cosmology, as 
for instance used by Hartle and Hawking in their semiclassical 
evaluation of the wave function of the universe \cite{hh}.
In Euclidean signature and proper-time coordinates, the
metrics are of the form 
\beq\label{3.1}
ds^2 = dt^2 + a^2(t) d\Omega_3^2,
\eeq
where the scale factor $a(t)$ is the only dynamical variable
and $d\Omega_3^2$ denotes the metric on the three-sphere.
The corresponding Einstein-Hilbert action is
\beq\label{3.2}
S^{\it eff}= \frac{1}{G_N} \int \d t \left( -a(t)
\left(\frac{\d a(t)}{\d t}\right)^2
-a(t) + \lam a^3 (t)\right).
\eeq
If no four-volume constraint is imposed, $\lambda$ is the
cosmological constant. If the four-volume is fixed
to $V_4$, such that the discussion parallels the
computer simulations reported above,
$\lam$ should be viewed as a Lagrange multiplier
enforcing a given size of the universe. In the latter
case we obtain the same effective action as that extracted
from the Monte Carlo simulations in \rf{2.11},
{\it up to an overall sign}, due to the infamous
conformal divergence of the classical Einstein action evident in
\rf{3.2}. From the point of view of the {\it classical} equations of
motion this overall sign plays of course no role.
Let us compare the two potentials relevant for the
calculation of semiclassical Euclidean solutions associated
with the actions \rf{3.2} and \rf{2.11}. The
``potential''\footnote{To obtain a standard potential -- without changing
``time" -- one should
first transform to a variable $x=a^{\frac{3}{2}}$ for which
the kinetic term in the actions assumes the standard quadratic
form. It is the resulting
potential $\tilde V(x)=-x^{2/3}+ \lam x^2$ which in the case of
\rf{2.11} should be modified
for small $x$ such that $\tilde V'(0)= 0$.} is
\beq\label{3.3}
V(a)= -a + \lam a^3,
\eeq
and is shown in Fig.\ \ref{fig4}, without and with small-$a$ modification,
for the standard
minisuperspace model and our effective model, respectively.
The quantum-induced difference for small $a$ is important
since the action  \rf{2.11} admits a classically stable solution $a(t)=0$
which explains the ``stalk'' observed in the computer simulations (see Fig.\ \ref{fig0}).
Moreover,  it is appropriate to speak of a Euclidean ``bounce'' because
$a=0$ is a local maximum. If one therefore
{\it naively}  turns the potential upside down
when rotating back to Lorentzian signature, the metastable state
$a(t)=0$ can tunnel to a state where $a(t) \sim V_4^{1/4}$,
with a probability
amplitude per unit time which is (the exponential of) the Euclidean
action.

In order to understand how well the semiclassical action
\rf{2.11} can reproduce the Monte Carlo data, that is, the correlator
\rf{v-v} of Fig.\ \ref{fig2}, we have solved for the
semiclassical bounce using \rf{2.11}, and presented the result
as the  black curve in Fig.\ \ref{fig2}.
The agreement with the real data generated by the
Monte Carlo simulations is clearly perfect.

The picture emerging from the above for the effective dynamics
of the scale factor resembles that of a universe created by
tunneling from nothing (see, for example, \cite{vilenkin,linde,
rubakov}), although the presence of a preferred notion of
time makes our situation closer to conventional quantum
mechanics. In the set-up analyzed here, there is apparently a
state of vanishing spatial extension which can ``tunnel" to
a universe of finite linear extension of order $a \sim V_4^{1/4}$.
Adopting such a tunneling interpretation, the action
of the bounce is
\beq\label{4.1}
S_{V_4}^{\it eff} \sim \frac{V_4^{1/2}}{G_N},
\eeq
and the associated probability per unit proper
time for the tunneling given by
\beq\label{4.2}
P(V_4) \sim e^{-S^{\it eff}_{V_4}}.
\eeq

\section{Discussion}

Causal dynamical triangulations (CDT) provide a regularized model 
of quantum gravity, which uses a class of piecewise linear
geometries of Lorentzian signature (made from flat triangular 
building blocks) to define the regularized sum over geometries.
The model is background-independent and has a diffeomorphism-invariant 
cut-off.  For certain values of the bare gravitational and 
cosmological coupling constants 
we have found evidence that a continuum limit exists. 
The limit has been analyzed by rotating the sum over geometries 
to Euclidean signature, made possible by our 
use of piecewise linear geometries. The geometries included in the 
sum thus originate from Lorentzian-signature spacetimes,
a class different from (and smaller than) the class of geometries one 
would naturally include in a ``native" Euclidean path integral.
We have concentrated on computing a particular 
diffeomorphism-invariant quantity, the proper-time propagator, 
representing the 
sum over all geometries whose space-like boundaries are separated by a 
geodesic distance $T$. The sum over such geometries allows a
simple and transparent implementation in terms of the 
above-mentioned building blocks.

In the Euclidean sector of the model, which can be probed by computer 
simulations we observe a four-dimensional macroscopic universe that 
can be viewed as a ``bounce''. 
When we integrate out ({\it after} having constructed the full path 
integral) all geometric degrees of freedom
except for the global scale factor, the large-scale 
structure of the universe (the bounce) is 
described by the classical general-relativistic solution
for a homogenous, isotropic 
universe with a cosmological constant on which (small) quantum fluctuations
are superimposed. We find this result remarkable in view of the 
difficulties -- prior to the introduction of causal dynamical triangulations --
to obtain dynamically any kind of ``quantum geometry" resembling 
a four-dimensional universe.  
In our construction, the restrictions imposed by causality before rotating 
to Euclidean signature clearly have played a pivotal role.

A number of issues are being addressed currently to obtain a more complete
understanding of the physical and geometric properties of the
quantum gravity theory generated by CDT, and to
verify in more detail that its classical limit is well defined.
Among them are:
%\vspace{6pt}
\\
\noindent (i) A better understanding of the renormalization of the 
bare coupling constants in the continuum limit, with the currently
favoured scenario being that of asymptotic safety \cite{weinberg}. 
There are very encouraging agreements between the results of 
CDT and those of a Euclidean renormalization group approach 
\cite{reuteretc} (see \cite{kawai} for older, related work). 
In particular, both approaches obtain a scale-dependent 
spectral dimension which varies between four on large and 
two on short scales.
%\vspace{6pt}
\\
\noindent  (ii) An identification and measurement 
of the ``transverse'' gravitational degrees of freedom, to complement
the information extracted so far for the scale factor only. 
For background-independent and coordinate-free formulations like CDT
we still lack a simple and robust prescription for how to 
extract information about the transverse degrees of freedom,
a quantity analogous to the Wilson loop in non-abelian gauge
theories. 
%\vspace{6pt}
\\ 
\noindent (iii) The inclusion of matter fields in the computer simulations.
Of particular interest would be a scalar field, playing the role of an
inflaton field. While it is straightforward to include a scalar field in the 
formalism, it is less obvious which observables one should measure, 
being confined to the Euclidean sector of the theory. 
Based on a well-defined CDT model for the nonperturbative 
quantum excitations of geometry and matter,
moving the discussion of quantum
cosmology and various types of inflation 
from handwaving arguments into the realm of quantitative analysis
would be highly desirable and quite possibly already within reach.

\subsection*{Acknowledgment}
All authors acknowledge support by
ENRAGE (European Network on 
Random Geometry), a Marie Curie Research Training Network in the 
European Community's Sixth Framework Programme, network contract
MRTN-CT-2004-005616. R.L.\ acknowledges support by the 
Netherlands Organisation for Scientific Research (NWO) 
under their VICI program.
J.J.\ was partly supported by the Polish Ministry of Science and Information
Society Technologies grant 1P03B04029(2005-2008).

\end{document}